# INTEROPERABILITY, TRUST BASED INFORMATION SHARING PROTOCOL AND SECURITY: DIGITAL GOVERNMENT KEY ISSUES

Md.Headayetullah[1] and G.K. Pradhan[2]

[1]Department of Computer Science and Engineering, SOAU, Bhubaneswar, India
headayetullahphd@gmail.com
[2]Department of Computer Science and Engineering, SOAU, Bhubaneswar, India
gopa_pradhan@yahoo.com

### ABSTRACT

*Improved interoperability between public and private organizations is of key significance to make digital government newest triumphant. Digital Government interoperability, information sharing protocol and security are measured the key issue for achieving a refined stage of digital government. Flawless interoperability is essential to share the information between diverse and merely dispersed organisations in several network environments by using computer based tools. Digital government must ensure security for its information systems, including computers and networks for providing better service to the citizens. Governments around the world are increasingly revolving to information sharing and integration for solving problems in programs and policy areas. Evils of global worry such as syndrome discovery and manage, terror campaign, immigration and border control, prohibited drug trafficking, and more demand information sharing, harmonization and cooperation amid government agencies within a country and across national borders. A number of daunting challenges survive to the progress of an efficient information sharing protocol. A secure and trusted information-sharing protocol is required to enable users to interact and share information easily and perfectly across many diverse networks and databases globally. This article presents (1) literature review of digital government security and interoperability and, (2) key research issue trust based information sharing protocol for seamless interoperability among diverse government organizations or agencies around the world. While trust-based information access is well studied in the literature, presented secure information sharing technologies and protocols cannot offer enough incentives for government agencies to share information amid them without harming their own national interest. To overcome the drawbacks of the exiting technology, an innovative and proficient trust-based security protocol is proposed in this article for sharing of top secret information amid government intelligence agencies globally. The trust protocol intended assures the enhanced interoperability of any modern digital government by sharing secure and updated information among government intelligence agencies to avoid any threatening deeds.*

## 1. INTRODUCTION

Government is a prime stasher and supplier of data and information, provider of information based services and client of information technologies [1]. Digital government can be defined as the "civil and political conduct of government using information and communication technologies" [2]. This includes the provisioning of services and the management of governmental processes. Such technologies can make powerful nation with greater access to services and more flexible and effective means of participating in government, foremost to enhanced citizen-government communication and consequently an overall development in society. These interests must be reasonable, however, with both security and privacy concerns. Two separate hitherto interconnected goals that are vital to the provisioning of digital government are security and privacy. Privacy is a social provision that is well thought-out wanted in many conditions. Security mechanisms make possible privacy. This message examines preferred issues concerning to privacy and security in digital government, in exacting it discusses the policy shift away from the conventional sight of non-repudiation that has been brought about by technical authentication mechanisms and developments in security policy while September 11th, 2001[36].

Enhanced interoperability amid public organizations and between public and private organizations is of vital significance to make digital government added victorious [37, 49]. Interoperability in digital government is well studied in the literature [38, 39, 40] the interoperations, which are used to illustrate among technical system, societal, political, and organizational. The recruitment of electronic information across organizations has the prospective to renovate and improve information interactions. The recent information exchange is, thus far, regularly ineffective and error-prone. Exchanges of information and services are disjointed and intricate, overwhelmed by technical and organizational troubles. In support of digital government to be booming it must expand responsive, citizen centric, liable, apparent, successful and proficient government operations and services. The amalgamation of government information resources and processes, and thus the interoperation of autonomous information systems, are crucial to accomplish these goals. However, mainly integration and interoperation pains mug solemn challenges and restrictions. The widespread application of information and communications technology (ICT) for the delivery of government services is the key next step in reinventing government, in a nutshell, encouragement digital government [8]. Digital (or electronic, or online; or connected) government fosters the use of information and technology to aid and enhance public policies and government operations, employ citizens, and offer widespread and well-timed government services. In the other words, digital government is producing a secure, apparent, and economical communication among government and citizens, government and business enterprises and relationship amid governments [50]. Interoperability is the capability of government organizations to share information and amalgamate information and business processes by means of widespread principles and work practices [41].Interoperability refers to a possessions of miscellaneous systems and organizations which enables them to work collectively[42, 43]. The initiatives taken for digital government are difficult alter efforts proposed to make use of novel and rising technologies to aid transformations in the operation and effectiveness of government [12].

Digital government, intended at optimizing its interior and exterior functions with the sustain of Information Technology (IT), offers a set of tools to the government, the citizen and business that can prospectively transform the means of interaction, service delivery, knowledge utilization, policy development and implementation, participation of citizens in governance, and reforms in public administration and good governance goals are fulfilled [13]. Information is the key resources of the Government around the world. It is opting to information sharing as a tactic for maximizing the value of information in providing services and responding to problems [21, 26]. On the appearance of collaboration and information sharing across domains as key elements of success, the potential for successful government Information Technology (IT) innovations is primarily important [3]. Information sharing illustrates diverse things to diverse government sectors at different periods of time. The information can be collection and sharing of intelligence amid two security divisions, or sharing actual e-crime data, data observations, notes of surveillance, scientific facts, commercial transaction data, and other. While there is a lack of accessibility of standard methods for e-government information sharing, the modes of information sharing is presently not monitored, authentic and recorded recurrently [23, 27]. Electronic government (or Digital Government) refers to the deliverance of government services (information, interaction and transaction) through the use of information technology (IT), a division can be made between the front and back offices of public service deliverance organizations. The transaction between citizens and social servants occurs in the front office, while registration and other activities take place in the back office. It is established that back-office support is a grim blockage in e-government due to diverse interoperability problems. One basic exploit to develop information sharing is standardization in information systems [45]. Interoperability of systems enables interoperability of organizations. Systems interoperability is concerned with the capability of two or more systems or components to exchange information and to use the information that has been exchanged. Organizational interoperability is concerned with the ability of two or more units to provide services to and accept services from other units, and to use the services so exchanged to allow them to manage efficiently collectively [46]. Semantic interoperability is fraction of the interoperability challenge for networked organizations. Inter-organizational information systems merely work when they communicate with other systems and interact with citizens. This aspect of interoperability can only be met if communication principles are applied. A standards-based technology proposal allows partners to implement a conventional business function in a digitally superior way. A widespread information systems platform, afterward, is a set of principles that allows network participants to communicate and accomplish business processes electronically. A difference should be made between interoperability and integration. Integration is the forming of a superior unit of government entities, provisional or permanent, for the intention of merging processes along with sharing information. Interoperation in e-Government occurs while self-governing or diverse information systems or their components controlled by different jurisdictions, administrations, or exterior associates work collectively in a predefined and established ahead approach [47].

A number of research challenges subsist in information systems that offer international collaborations amid governments: information management across agencies and organizations, transparent interoperation across diverse information networks, and share of multilingual information [11, 12, 28] and the sharing of information is not for all time confident to be risk free. It may comprise illicit access, malicious alteration, and damage of information or misinformation, computer intrusions, copyright infringement, privacy violations, human rights violations and more. One of the most vital issues for development of efficient e-government architecture is secure sharing of information amid diverse government agencies. Government agencies visage a number of multifaceted worldwide problems like: border control, illegal immigration, terrorism, and bio-security threats. The multifaceted global problems could be solved by efficient collaboration and information sharing amid the agencies [4]. Information sharing is central to enhance the security of the country and is a vital factor in rising entire and sensible approaches for protection alongside looming terrorist attacks. Terrorism is definitely one of the significant problems all over the world [18], and it is ascertained that an effective and secure information sharing system amid global intelligence agencies will aid a more strict control over terrorism. There have been so many occurrences of terrorist activities witnessed in world that would have been prevented by effective trust based information sharing.

Terrorism has attained serious extent behind the twin towers attack on September 11 at the World Trade Centre in United States of America (USA). The whole world witnessed the full overfed pictures that showed the unexpected vertical fall down of the commercial might of USA. The September 11 attack and the follow investigation confirmed the survival of a serious information sharing problem amid the relevant federal government agencies, and the problem could cause generous deficiencies in terrorism attack detection [14]. On 26 November, 2008, the world experienced another most exposed sudden disaster, which was an outburst of unsociable activity beside common people of India, where more than a couple of hundred were dead and several hundreds were wounded [20]. The above are some critical catastrophes that made the authorities of worldwide intelligence agencies repeatedly assert the need for a more effective information sharing system. A secure and trusted information-sharing atmosphere is a requirement to permit government agencies to cooperate with and share information easily and faultlessly crosswise several diverse networks and databases nationally [5,13]. Building a wide basis for information sharing necessitates trust among all information sharing associates. The panic that shared information will not be protected effectively or used properly and that sharing will not always occur reciprocally are reason for lack of trust [14].That means there are different interoperability among the government agencies. Collection and sharing of information of some organization is affected by confidentiality concerns [44].Formerly, sharing of information among law enforcement agencies has occurred in a very top secret way normally, only by individual to individual or case by case basis. Effective information sharing amongst different communities at diverse levels of government – national, state, regional, and local – has developed into a height priority of world governments, whose leaders frequently proclaim the need for more effective information sharing to boost homeland security efforts [14].

The modern information sharing technologies can be categorized into two types: (i) privacy-preserving information sharing, wherever two communicating parties with information m and n respectively communicate with each other such that a function of m and n, symbolized f(m, n) is computed and shared by the two parties, devoid of illuminating the privacy of m and n and (ii) non-privacy-preserving information sharing, wherever two communicating parties with information m and n correspondingly share (part of) m and/or (part of) n along with f(m, n) [29, 30].The capability to share terrorism-related information can perform a unification of the federal, state, and local government agencies efforts, as well as the private sector in prevention or sinking terrorist attacks. Countless government agencies have lack of effective interoperability as well as interoperation among them and neglect to share information mainly owing to a) internal conflicts between them b) horror of harming their own national welfare and c) most notably the terror of information being hacked. Information sharing amongst government agencies necessitates a distinct, more preventive trust model primarily due to two reasons: (1) Availability of highly responsive information (2) the requirement of a more liable and fair information sharing procedure to conquer the differences and conflicts-of-interest existing between agencies [14]. The above entire factors demand the need for an effective interoperability among the agencies and which needs for a trust based effective information sharing system. Even though trust-based information access is studied well in the literature [19, 15, 16, 17, 7] the available trust models, which are on the core of specialized attributes, could not offer effectual information sharing amid government agencies. This paper presents an innovative and well-organized trust based security protocol for providing top secret information sharing among worldwide government intelligence agencies devoid of impairing their individual national safety. The demand for confidentiality may be a function of legal requirements for managing citizen data that are of a personal nature or of the handling of data whose disclosure can in some way threaten the operation or security of citizens or government. In our earlier research we have presented an efficient and secure information sharing approach for security personnel's to share information among them and with security related Government departments. The proposed approach is mainly adapted to suit the subsequent set-up. Consider, for instance, a local law enforcement officer at a tollgate by the side of a landmark. The benchmark approach for traffic control necessitates the officer to appeal and confirm the individual's driving license and vehicle registration. However, the law enforcement officer could also desire to ensure with an extensive range of other computer applications, such as immigration databases, criminal information and intelligence repositories, and counter-drug intelligence databases that may be owned by outside organizations, for example, Central Bureau of Investigation, the Drug Enforcement Administration, and the Department of Homeland Security. The precision and the amount of information shared between communicating government intelligence departments is based on the predefined rank of the security personnels maintained between the communicating government departments. The proposed role and co-operation based approach achieves data integrity using MD5 Algorithm, confidentiality and authentication using public key infrastructure and department verification using a mapping function[51].

This work is an improved edition of our previous research [35], which improves the security of the existing trust-based security protocol by providing authentication, for confidential exchange of top secret information amongst worldwide government intelligence agencies devoid of harming their own national interests. The present work includes the interoperability and how it is key enabler to Digital Government. Inter-organizational information integration has become a key enabler for Digital Government. Integrating and sharing information crosswise conventional government restrictions involves multifaceted communications between a varieties of participants all using complex practical and governmental processes. The proposed article is well studied about digital government interoperability, secure information sharing protocol and information integration, which are key enablers of successful modern digital government. This paper also presents effective digital government interoperability by sharing and integrating the top secret information among worldwide government intelligence agencies without harming their national interest. In general, digital government philosophy is on the basis of the implication that various government agencies are organized to assist and share findings throughout a network infrastructure. A secure means to support information transfer fosters the daring and teamwork among the government intelligence or security personnel agencies. By the assist of the designed trust-based security protocol, the government intelligent agencies can effectively share information about terrorists and their misapprehension in a protected approach. The credibility of information shared is based on the trust level maintained among communicating government intelligence agencies. The proposed article covers three section of modern digital government research, such as, normative, perspective and evaluative by making use of basic concept of interoperability, security and information sharing protocol, advance technological approach and to offer option for general user to make query to the concerned department about terrorist and their activities respectively.

The proposed security protocol provides data integrity using MD5 Algorithm, confidentiality using public key cryptosystem and password verification system, controlled privacy using trust level and agency verification using a mapping function. The rest of the paper is well thought-out as follows: A concise review of the researches related to the digital government interoperability, significance of trust-based protocols for secure information sharing among communicating parties such as government intelligence agencies or security personnel is offered in part 2. The proposed trust-based protocol for successful and secure information sharing is offered in part 3. Part 4 discusses the experimental results obtained and finally the conclusions are summed up in part 5.

## 2. REVIEW OF PREVIOUS RESEARCH

A copious of trust based information sharing protocols has been proposed by researchers for effective information sharing between communicating parties and to set up effective interoperability among the digital government agencies. Digital government interoperability can be achieved by sharing and integrating top secret information among the communicating parties. The developments of trust-based secure information sharing protocols and to set up proper interoperability among communicating parties are the leading research areas in digital (or electronic) government. Now, we present a brief review of special noteworthy contributions from the existing literature.

Papazoglou et al. [47] have proposed interoperability standards. The propose standards necessitate consistency in four dimensions such as: (i) technology, (ii) syntax, (iii) semantics, and (iv) pragmatics. Technology standards concern middleware, network protocols, security protocols, and the resembling. Syntax standardization means that the network organization has to concur on how to incorporate diverse applications based on the structure or language of the messages exchanged. Generally, suitable data structures are elected to signify eminent constructs (for example, invoice descriptions).Semantic standards comprise agreements in extension to syntactic agreements on the meanings of the terms used for an enterprise's information systems. Pragmatic standards are agreements on practices and protocols triggered by specific messages, such as orders and delivery notifications. In addition, they proposed novel e-business models to reduce costs and improvement of digital government internal and external operations foremost services. This proposed model supports citizen-centric services by integrating and sharing information among the stakeholders both vertically and horizontally.

Pardo et al. [37] have presented digital government concept as well information sharing process. According to Pardo, digital government, electronic government, and electronic governance are terms that have become identical with the use of information and communication technologies (ICT) in government agencies. Inter-organizational information assimilation has become a key enabler for digital government. The proposed article described two complex information sharing process such as: practical process and governmental process. In practical process the system designers and developers must regularly exploit over dilemma related to the survival of numerous platforms, varied database designs and data structures, extremely variable data quality, and irreconcilable network infrastructure. In governmental view, these practical processes regularly employ novel work processes, mobilization of restricted assets, and developing inter-organizational interaction. The proposed articles present the necessary changes in societal communication for digital government information sharing process by taking into account of group decision–making, learning, perceptive, conflict resolution and trust building among government agencies. Trust based information sharing is more important in modern digital government process.

Scholl et al. [40] have presented electronic government (or digital government) interoperability and nine constraints which influences on electronic government integration and interoperability. Electronic (or digital) government interoperability is the technological ability for electronic (or digital) government interoperation. The proposed article described the nine constraints as follows: (a) Constitutional/legal constraints: Integration and interoperation may be out-and-out illegal since the democratic constitution requires powers to be alienated into separate levels and branches of government. The US constitution, for example, separates government into federal, state, and local government levels, and into legislative, judicial, and executive branches. Entire integration and interoperability between and among branches and levels might upset constitutional checks and balances. In contrast, the constitution also affords and sanctions integration and interoperation within assured restrictions. (b) Jurisdictional constraints: While under the constitution, governmental and non-governmental constituencies operate autonomously from each other and own their information and business processes, integration, interoperation, and information sharing cannot be imposed on them. Moderately, as an independent entity, each constituency's contribution in any interaction is unpaid. Yet, using jurisdictional authority, the government entity can employ in integration and interoperation with other entities. (c) Collaborative constraints: Organizations are diverse in terms of their character and willingness for cooperation and interoperation with others. Past familiarity, socio-political organization, and leadership style influence the degree of willingness and proficiency of potential interoperation. In cases of complementary leadership styles, adequate socio-political organization, and encouraging record skill, integration and interoperation might flourish.(d) Organizational constraints: In this constraints the integration and interoperation might difficult to establish as organizational processes and resources may differ among organization rapidly. Still, it facilitates improved degrees of integration and interoperation while the organization supports its organizational view. (e) Informational constraints: Transactional information might be more freely shared compared to strategic and organizational information. Information quality standards arise when information are collecting, integrating and sharing across several domains. At rest, information stewardship fosters the use of shared information, which in turn fosters stewardship for giving out information. (f) Managerial constraints: Interoperation becomes really extra intricate the added parties with diverse welfare and wishes become apprehensive. Therefore, the needs of the related management task might exceed the management ability of interoperating partners. However, beside the ranks of shared interests, interoperation and integration can embarrass. (g) Cost constraints: Integration and interoperation between varied constituencies might be limited to the lowest frequent denominator in terms of availability of funds: likewise, unexpected budget constraints might cover-up serious challenges to long-term inter-operation projects ultimately. Equally, information-sharing initiatives have evidently helped control costs. In the cost margins of the relevant partners, certain projects appear to be sustainable. (h) Technological constraints: The heterogeneity of digital government policy and network capabilities might limit the interoperation of systems to relatively low standard. On the contrary, a growing number of digital government information systems might attach to superior standards ultimately, such that superior interoperation becomes possible. (i) Performance constraints: While performance checks warn, the higher the number of interoperating partners, the lower is the generally system performance in provisions of response time. So far, the highlight on prioritized needs might assist less but more successful interoperations. The presented article proposes how to eliminate these nine limitations of digital (or electronic) government interoperation. This proposed work will help to attain digital (or electronic) government operations and services that are efficient, agile, citizen-centric, accountable, apparent and efficient. The integration of government information resources and processes, so the interoperation of autonomous information systems, is crucial to achieve these objective. Conversely, mainly integration and interoperation efforts mug serious challenges and limits.

Chen et al. [48] have presented stages of maturity models of electronic (or digital) government. The maturity model of electronic(or digital) government provides the users information and services in great scale of density across numerous dimensions of electronic(or digital) government .These models propose that electronic(or digital) government capabilities begin moderately and chiefly provide fixed, one-way information, but grow progressively more refined and add interactive and transactional capabilities. These models envisage an imperative progression of electronic (or digital) government that includes horizontal and vertical integration and the development of true portals and flawless inter-organizational exchanges. Vertical and horizontal integration are the key facet of digital (electronic) government research. There were three different types of electronic( or digital) government maturity models such as: (a) First model displays, a few aspect, the policy, technology, data, and organizational concerns that must be determined for organizations to growth to higher levels of electronic (or digital) government maturity with a helper hoist in benefits for equally government organizations and end-users. It is necessary to have both higher levels technology and organizational density to achieve more established levels of electronic (or digital) government. (b) Second model identified the following four stages of electronic (or digital) government integration: (1) catalogue with online presence, catalogue presentation, and downloadable forms, (2) transaction with services and forms online, working database, and supporting online transactions, (3) vertical integration with local systems linked to higher levels systems and within related functionalities, and (4) horizontal integration with systems integrated across different functions and real one-stop shopping for citizens. (c) Third model stresses intensifying levels of data integration required for proper transformational electronic (or digital) government, but warns that such data integration raises significant confidentiality concerns when the data involves personally identifiable information. It was commented that these models entail, but only sometimes make explicit, that the complexity of these diverse forms of integration have likely resulted in countless organizations triumph the highest level of electronic (or digital) government maturity. Semantic interoperability is defined as the level to which information systems using diverse expressions are able to converse. Organizational interoperability is defined as the level to which organizations using diverse effort practices are able to converse. The proposed article discussed the stage-of- growth model for interoperability in electronic (or digital) government and how it assists to make successful digital (or electronic) government. The presented article is useful for scholars and practitioners to identify the present improvement stages for interoperability in digital government and aid to prepare for upcoming improvements in digital government research. Based on the reviewed literature on systems interoperability and stages-of-growth models, we are now ready to state that a prospective maturity model for interoperability in electronic (or digital) government will aid to share proper information among the different organization such as government sector, public sector and private sector. Thus, information sharing protocol for effective interoperability is one the key issue in digital government research and we need to review sufficient previous research articles to set up effective information sharing protocol.

Peng Liu et al. [14] have proposed a novel interest-based trust model and an information sharing protocol to triumph the problem of information sharing between government agencies. The proposed protocol integrated a family of information sharing policies, along with information exchange and trust negotiation, interleaved and interdependent upon each other. Moreover, the protocol was implemented by making use of the rising technology of XML Web Services. The implementation was completely well-suited with the Federal Enterprise Architecture (FEA) reference models and can be integrated explicitly into presented electronic (or digital) government systems.

Jing Fan et al. [25] have presented a conceptual model for information exchange in electronic (or digital) government infrastructure. They figured out that the Government-Government (G2G) information sharing model will aid in contribution an understanding for G2G information sharing and will assist decision makers in framing decisions with regards to contribution in G2G information sharing. They tested the proposed conceptual model with the purpose of discovering the factors persuading the partaking in electronic (or digital) government information sharing and emphasizing the conceptual model via case study under Chinese government system.

Fillia Makedon et al. [23] have presented a negotiation-based sharing system called SCENS: Secure Content Exchange Negotiation System developed at Dartmouth College with the aid of many interdisciplinary experts. SCENS was a multilayer scaleable system that brings about surety to transaction safety via numerous security mechanisms. It was based on a metadata description of heterogeneous information and was applied to a number of diverse domains. They demonstrated that with susceptible and distributed information the government users might bring about an agreement on the conditions of sharing information by way of negotiation.

Xin L. [22] has set up a distributed information sharing model and also ponders the technique standard support of the model. It was deduced that the cost of managing the government information exchange and cooperation between agencies will be decremented with an augment in the ability and efficiency of agencies' collaboration owing to the secure electronic (or digital) government information sharing solutions.

Nabil R. Adam et al. [32] comprise thought-out on confronts in integration, aggregation and secure sharing of information for offering situation awareness and response at the strategic level. The proposed system based on context-sensitive parameters, filters, integrates, and proficiently visualizes the data extracted from different autonomous systems essential to get a common operational picture. One remarkable confront found was to make certain secure information sharing. Information sharing remains to be a prime intricacy owing to the data privacy and ownership concerns and a wide range of security policies approved within diverse government agencies. Nabil Adam et al. [33] have presented a two tier RBAC approach to offer security and discriminative information sharing between virtual multi-agency response team (VMART) and as requirements arise it allows VMART expansion by admittance of new collaborators (government agencies or NGOs). They also offered a coordinator Web Service for each member agency. The coordinator Web Service workings with the subsequent responsibilities: authentication, information spreading, information acquisition, role creation and enforcement of predefined access control policies. Realization of Secure, selective and fine-grained information sharing was accomplished by the XML document encryption in compliance with analogous XML schema defined RBAC policies.

Achille Fokoue et al. [34] have set up logic for risk optimized information sharing by utilizing rich security metadata and semantic knowledge-base that encapsulates domain specific concepts and relationships. They established that the approach was: (i) flexible: such as, tactical information decay sensitivity in connection with space, time and external events, (ii) situation-aware: such as, encodes need-to-know based access control policies, and more notably (iii) supports explanations for non-shareability; these explanations with rich security metadata and domain ontology allows a sender to perceptively accomplish information transformation with the target of sharing the transformed information with the recipient. Still, they have offered a secure information sharing architecture making use of a publicly available hybrid semantic reasoner and also presented a number of enlightening examples that accentuates the advantages of the approach in comparison to traditional approaches.

Ravi Sandhu et al. [24] have projected the ways by which recent Trusted Computing (TC) technologies could assist secure information sharing, those not offered with pre-TC technology. They have produced the PEI framework such as policy, enforcement and implementation models, and set up its application in tentative the problem with synthesizing solutions for it. The framework made possible the detailed examination of potential TC applications for secure information sharing in their upcoming work.

Tryg Ager et al. [31] intended a set of policy-based technologies to easiness increased information sharing between government agencies without negotiating information security or individual privacy. The approach comprises: (1) fine-grained access controls that support deny and filter semantics for complex policy condition satisfaction; (2) a sticky policy ability that assists consolidation of information from multiple sources subject to the source's original disclosure policies of each; (3) a curation organization that permit agencies to apply and contrive item-level security classifications and disclosure policies; (4) an auditing system that accounts for the curation history of each information item; and (5) a provenance auditing method that traces information derivations over time to offer support in evaluating information quality. The eventual aspiration was to offer a scope to solve stupendous information sharing problems in government agencies and offer track for the expansion of upcoming government information systems.

## 3. INFORMATION SHARING PROTOCOL BASED ON TRUST

Government information is a major asset that must be maintained in trust and well managed by governments. A more importance has to be put forth by government institutions, at all levels, on the sharing of data and information between and amongst its trusted partners. Through the idea of meeting the growing demands and service expectations, information must be influenced and supported by coordinated, integrated solutions [6]. With handy information sharing solutions, government intelligence agencies will be competent to predict the security risks and attacks, including terrorist attacks. However, devising secure information sharing mechanism between government intelligence agencies is not trivial because they worry that their interests may be exposed when their information is shared with other agencies [22]. This section presents the proposed modern and expert trust-based security protocol for secure sharing of confidential information among government intelligence agencies.

The devised protocol is non privacy-preserving, but assures that both the source and the target agencies are ensured supreme confidentiality and authentication in information transfer. The intelligence agencies worry that the hacking of sensitive information shared would cause apprehension to their own national interests. This demands an efficient security protocol that offers confidential and authenticated information sharing with regards to the national interests of both the source and the target government agencies. In addition, there is probability that the target agency would misuse the secret information without the suitable approval of the source agency. The above case cannot be entirely averted in a non privacy-preserving protocol but could be restricted by availing information transfer based on the predefined trust level existing between the communicating government agencies. The government intelligence agencies make use of the devised protocol to share terrorist information in a secure manner. The credibility of information shared is based on the trust level maintained between communicating government intelligence agencies.

The proposed protocol ensures trust-based secure information exchange between communicating agencies. The prerequisites for the proposed trust-based security protocol includes: (1) The public keys of the communicating agencies; (2) A unique and complex mapping function. The communicating agencies achieve their public and private keys from a trusted Certificate Authority (CA). The predefined complex mapping function uniquely identifies the communicating agencies. The steps describing the proposed trust-based security protocol is organized in such a way that, the source agency first requests for some secret and valuable information followed by the corresponding target response based on the trust level maintained and a validation of the target response at the source agency. This propose articles works like a semi automated software agent who negotiate trust and share top secret information with each other on behalf of corresponding agencies. In this way human being partially relieved from efforts needed to run information sharing protocol and more efficient and timely information can be achieved. More timely information sharing can mean more timely detection of terrorist attacks and less damage caused by such attacks. In this articles system administrator acts as a software agent for the government intelligence agencies. The software agent works on duplicates database stored previously based on predefined trust (or privacy, or rank) of the government agencies. In addition, there is provision of general agencies (or users) to interact with intelligence agencies of the government to querying for information about unsocial, threatening, mistrustful activities of any terrorists organization. This mechanism mainly consists two agencies general agencies i.e. general user and intelligence agencies i.e. administrator. This protocol can be used as semi-automated software and simply user can request for registration to the administrator. Administrator verifies the user request and gives one password for further interaction for terrorist activities to avoid the premeditated activities of terrorist but there is limitations of the general user who can query for information about terrorist to the local government agencies and users have no right to change or modify any information. This protocol can use by intelligence agencies to share information from one government to another government in the worldwide on the basis of predefined trust. This article gives the provision of direct query system on the basis of previously stored information about terrorist and their activities. This research gives efficient way for each agency to share information with one another and sharing of information on the basis of direct query of the databases among the agencies. Database scheme information is based on trust and which is predefined among the agencies. For example, if agencies have already shared information with one another, then each agency would have already obtained significant scheme information about the other agency's databases. Such scheme information may facilitate both agencies to compose accurate, execute queries. By means of transfer a query directly from one agency to another, an agency will be able to identify the information on the basis of that the information will be executed. The proposed trust-based security protocol can established for Intelligence agencies follows:

## 3.1. STEPS IN THE PROTOCOL AT SOURCE AGENCY

### 3.1.1. STRUCTURING OF SOURCE REQUEST

The source intelligence agency requests for some secret information about terrorists and their mistrustful activities to the target intelligence agency. It is the job of the source intelligence agency to transmit the request in an incomprehensible probably encrypted way such that the hackers cannot take out any precious information or alter the information in the request. The frame of the source request involves the following stepladder:

1. A random number $R$ is designated and is encrypted by making use of the public key $K_S^{Pub}$ of the source agency. On target response, the encrypted random number $R_V$ will be used to certify that the response corresponds to the suitable source request. $R_V = Enc[R]_{K_S^{Pub}}$
2. A set of random values $S_R$, the source agency identifier and the request are shared with the encrypted random number $R_V$ to obtain $SE_{Data}$. The random values set $S_R$ will be utilized to authenticate the identity of the target. $S_R = \{r_1, r_2, r_3, \ldots, r_n\}$. The source encrypted data $SE_{Data}$ can be computed by making use the following equation.
$$SE_{Data} = R_V + src\ \ id + [S_R] + \mathrm{Re}\,quest$$
3. The MD5 Algorithm is engaged to compute the hash value $H_{val}$ of the $SE_{Data}$. Here MD5 algorithm is used to maintain the data integrity. $H_{val} = MD5[SE_{Data}]$
4. The hash value $H_{val}$, set of random values $S_R$, and the R*equest* are united and encrypted with the private key $K_S^{\mathrm{Pr}i}$ of the source agency to figure $SA_{Data}$. The encryption with the source private key reliably authenticates the source's request.
$$S_{Data} = S_R + \mathrm{Re}\,quest + H_{val}$$
$$SA_{Data} = Enc[S_{Data}]_{K_S^{\mathrm{Pr}i}}$$
5. The encrypted random number $R_V$ and the source agency identifier are then appended to $SA_{Data}$ to structure the user request $S_{\mathrm{Re}q}$. Ultimately, the user request is encrypted with the public key $K_T^{Pub}$ of the target agency to form $S_{\mathrm{Re}q}$
$$S_{\mathrm{Re}q} = Enc[R_V + src\ \ id + SA_{Data}]_{K_T^{Pub}}$$

The frame source agency request $S_{\mathrm{Re}q}$ possesses the encrypted random number $R_V$, the source agency identifier $src\ \ id$, $SA_{Data}$, all encrypted with the target's public key $K_T^{Pub}$. Currently, this source agency's request $S_{\mathrm{Re}q}$ is transmitted to the target intelligence agency.

### 3.2. STEPS IN THE PROTOCOL AT TARGET AGENCY

#### 3.2.1. VALIDATION OF THE SOURCE REQUEST

On receiving source agency's request, the target agency must verify source agency followed by validating the integrity of source agency's request. Next, on the basis of the trust level maintained with source agency, the target presents with an appropriate and top secret response to the source agency. The stepladder concerned in the integrity checking and authentication of source agency's request are as follows:

1. The request $S_{\mathrm{Re}q}$ is decrypted with the private key of the target $K_T^{pri}$. As the private key is the top secret property of the premeditated target agency, the target is secure that no one else can decrypt the request. $D_{Req} = Dec(S_{Req})_{K_T^{Pri}}$
2. The $D_{\mathrm{Re}q}$ obtained from step 1 contains $SA_{Data}$, $R_V$ and $src\ \ id$. The $SA_{Data}$ obtained is decrypted with the public key $K_S^{Pub}$ of the source agency. This authenticates that the request has originated from the claimed source agency. $D'_{\mathrm{Re}q} = Dec(SA_{Data})_{K_S^{Pub}}$ ; $D'_{\mathrm{Re}q}$ contains the set of random numbers $[S_R]$, the request and the hash value $H_{val}$.

3. The $SE_{Data} = (R_V + src\ id + [S_R] + \mathrm{Re}quest)$ is formed and its hash value $\overline{H_{val}}$ is computed with the help of the MD5 algorithm. Thus, $\overline{H_{val}} = MD5[SE_{Data}]$
4. If the hash value computed from the above step and the hash value present in the source agency's request are alike, it ensures that the request has not been tampered during data transmission.

$$if\ \ (H_{val} == \overline{H_{val}})\ \ then\ \ request\ is\ not\ tampered\ ;$$
$$end\ \ if$$

The decrypted source request contains the encrypted random number, the source agency identifier, the set of random values and the request. The target agency's response is on the basis of the trust level maintained between the source and target agency, which is maintained in a database. The trustworthiness and amount of information conveyed in the response depends on the trust level maintained with the communicating source agency.

### 3.2.2 STRUCTURING OF RESPONSE

The block diagram in Figure 1 illustrates the steps involved in structuring the target agency response based on trust level.

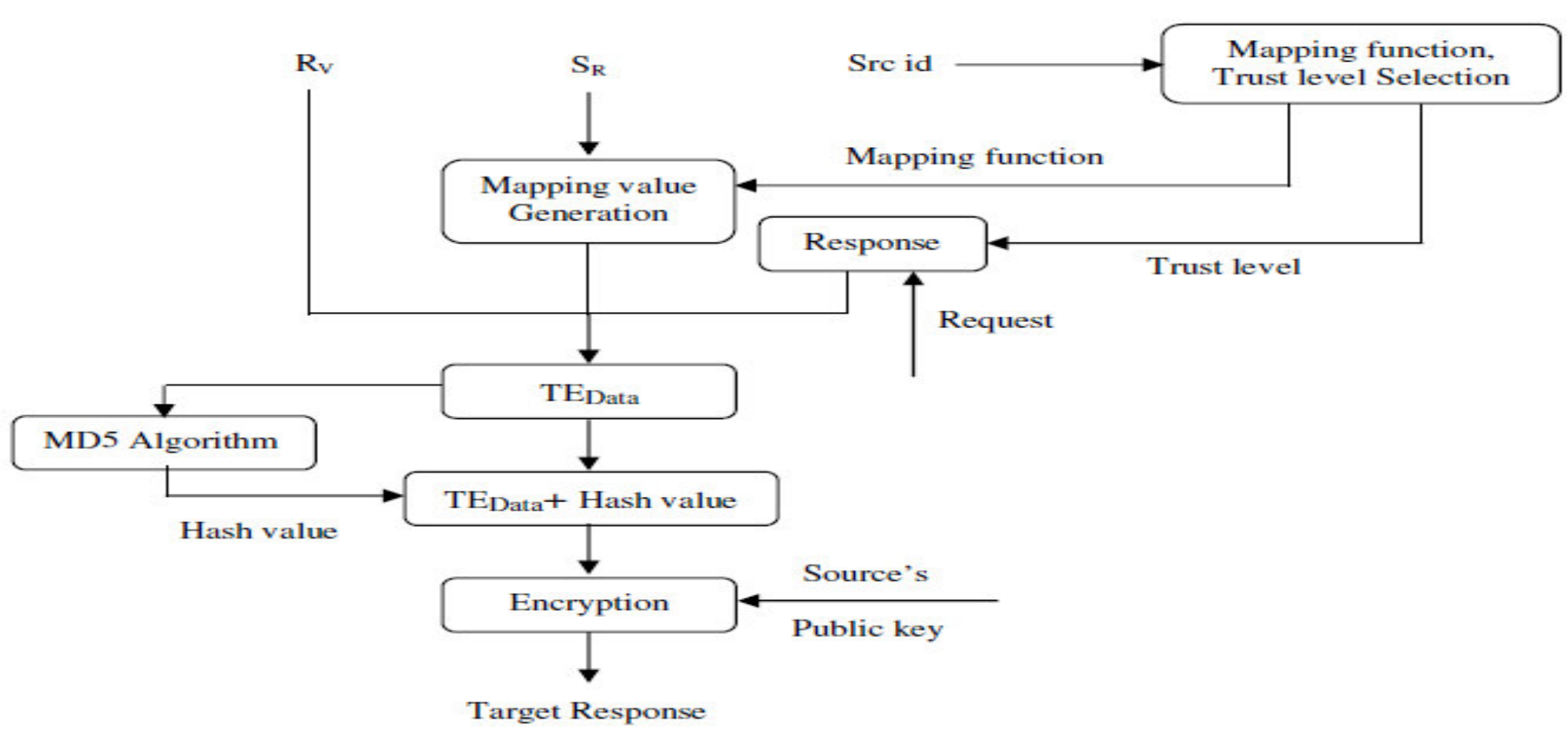

**Figure 1:** Structuring of response

The response to the corresponding source agency's request will be crafted as follows:

1. The target agency's database is scanned to reach the trust level maintained with the source agency. The trust level is a symbolization of the perceptive levels of their respective countries. The source agency identifier $(src\ id)$ serves as an index for the database search.
2. The encrypted random number $R_V$ in the source request is kept as such in the response.
3. A mapping function $M_{fn}$, uniquely defined between the communicating agencies is retrieved from the target database based on the source agency identifier. It is then applied to the set of random numbers in the source request to attain a mapping value $M_{val}$ . Next, its sine value is computed and denoted as $M'_{val}$. Thus, $M_{val} = M_{fn}(S_R)$

$M'_{val} = Sin(M_{val})$; Where $S_R = \{r_1, r_2, r_3, ......, r_n\}$, $M_{fn} = \{+,-,*,/\}$

4. The target agency determines the amount and credibility of confidential information to be shared with the source agency based on the trust level obtained from Step (1).
5. The equivalent response for the source request and the calculated mapping value are appended to the $TE_{Data}$. So, $TE_{Data} = [R_V + M'_{val} + \text{Re}\,sponse]$
6. The MD5 Algorithm is employed to compute the hash value of $TE_{Data}$ and is combined with $TE_{Data}$ to form the concluding response. $H_{val} = MD5[TE_{Data}]$
7. The structured concluding response of the target is finally encrypted with the public key of the source agency $K_S^{Pub}$ to obtain $T_{\text{Re}\,s}$. This ensures the confidentiality of the information shared. $T_{\text{Re}\,s} = Enc[TE_{Data} + H_{val}]_{K_S^{Pub}}$

Afterwards, the encrypted target response $T_{\text{Re}\,s}$ is sent back to the corresponding source agency.

## 3.3. STEPS IN THE PROTOCOL AT SOURCE AGENCY

### 3.3.1. VALIDATION OF TARGET RESPONSE

On getting the response from target, the source agency cannot believe it blindfold, but must make certain the following: (a) integrity of the target response (b) The response originated from the exact or intended target (Authentication) and (c) The response corresponds to the apt request of the source agency.

(1) The target response is decrypted using the private key of the source agency $K_S^{\text{Pr}\,i}$, which reveals the encrypted random number, mapping value, the response and the hash value.

$$ST_{\text{Re}\,s} = Dec\left(T_{\text{Re}\,s}\right)_{K_S^{\text{Pr}\,i}}$$

$$ST_{\text{Re}\,s} = [R_v + M'_{val} + \text{Re}\,sponse + H_{val}]$$

(2) The response is established for its integrity on the basis of the hash value computed using the MD5 algorithm

$$\overline{H_{val}} = MD5[R_V + M'_{val} + \text{Re}\,sponse]$$

$$if\ \ (H_{val} == \overline{H_{val}})\ then\ \ information\ \ is\ \ not\ tampered\ ;$$

$$end\ \ if$$

(3) The mapping value present in the response is recomputed at the source agency to ensure that the response came from the intended target.

$$if\ \ (M'_{val} == \overline{M'_{val}})\ then\ \ The\ \ target\ \ is\ valid\ \ ;$$

$$end\ \ if$$

(4) The encrypted random number in the target response is decrypted with the private key of the source agency $K_S^{\text{Pr}\,i}$ to guarantee if it is a valid response for the request made.

$$if\ (\text{R} == \text{Dec}[\text{R}_\text{V}]_{K_S^{\text{Pr}\,i}})\ \ The\ response\ is\ valid\ ;$$

$$end\ if$$

(5)After evaluating all these parameters, the source agency deems it as valid response from the target.

All the above steps assure that the proposed trust-based security protocol is effective in providing confidential, authenticated and secure information sharing. Advance communications between the agencies follow the procedures discussed above. The block diagram in Figure 2 shows the steps involved in the validation of the response

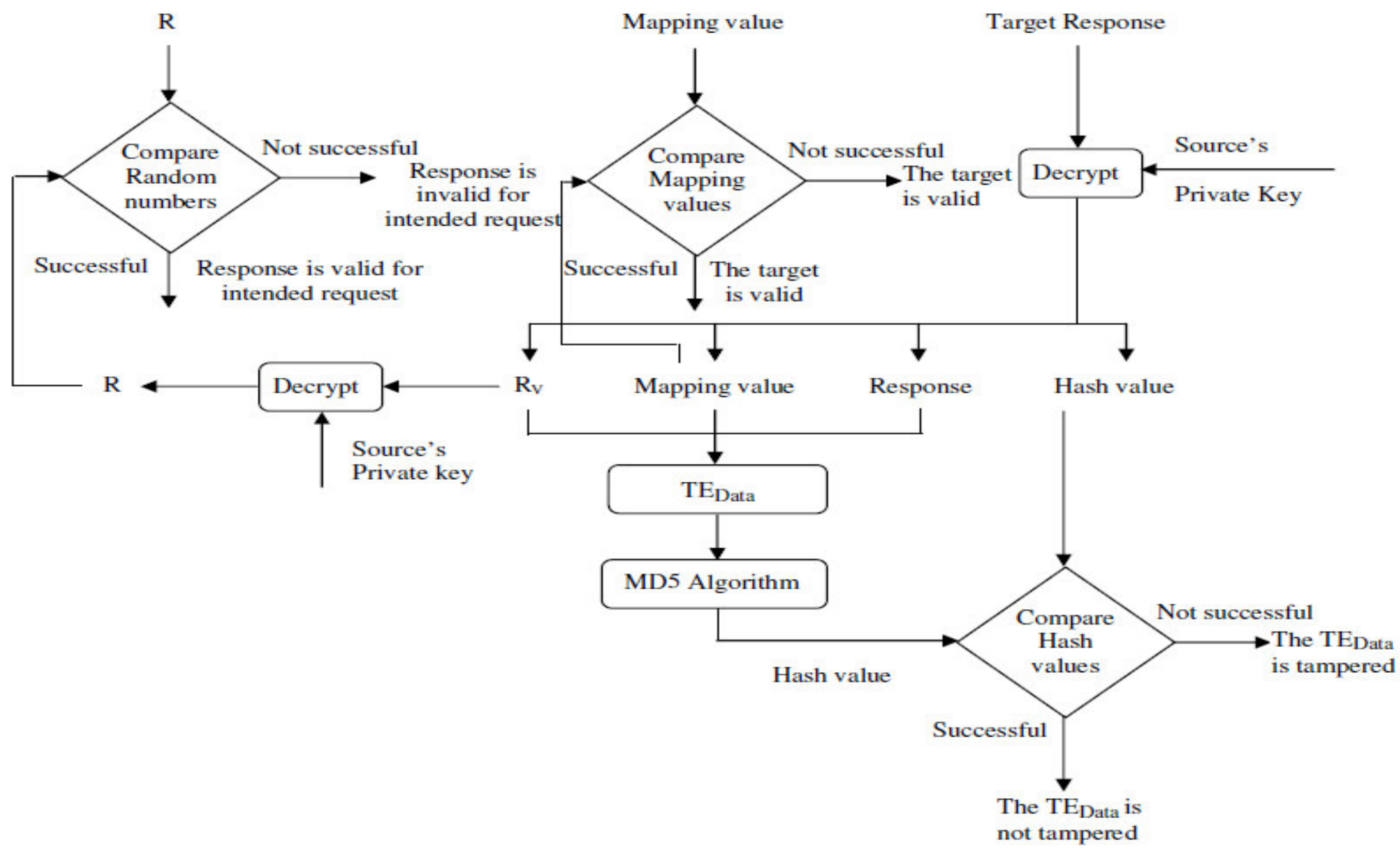

**Figure 2:** Validation of the target's response

## 4. EXPERIMENTAL RESULTS

The proposed innovative trust-based information sharing protocol is programmed in Java (JDK 1.6, Apache Tomcat 5.5 and MySQL 5.1). The obtainable experiment comprises: principally two levels: (a) Administrator level, acts as intelligence agencies which have central responsibilities of the whole database information system of governmental organization. (b) User Level, acts as general users, having provision to create an account to the authorised the government agencies to query for database information about terrorist activities. The general user has no authority to change or modification of database information about terrorists and interaction between government agencies and general user is based on trust. The proposed work will give the provision for general user to act together with the government intelligence agencies to protect the terrified activities of the terrorist such as Mumbai attack or the twin towers attack on September 11 at the World Trade Centre in United States of America (USA). The proposed work can be used as automated software or website with assist of web server. The government intelligence agency can also work together with general public of any country on the basis of trust or limited privacy with the aid of automated secure web site. The presented work illustrates some snapshots of our innovative work and which covers evaluative aspect of digital government research as shown in the below figures.

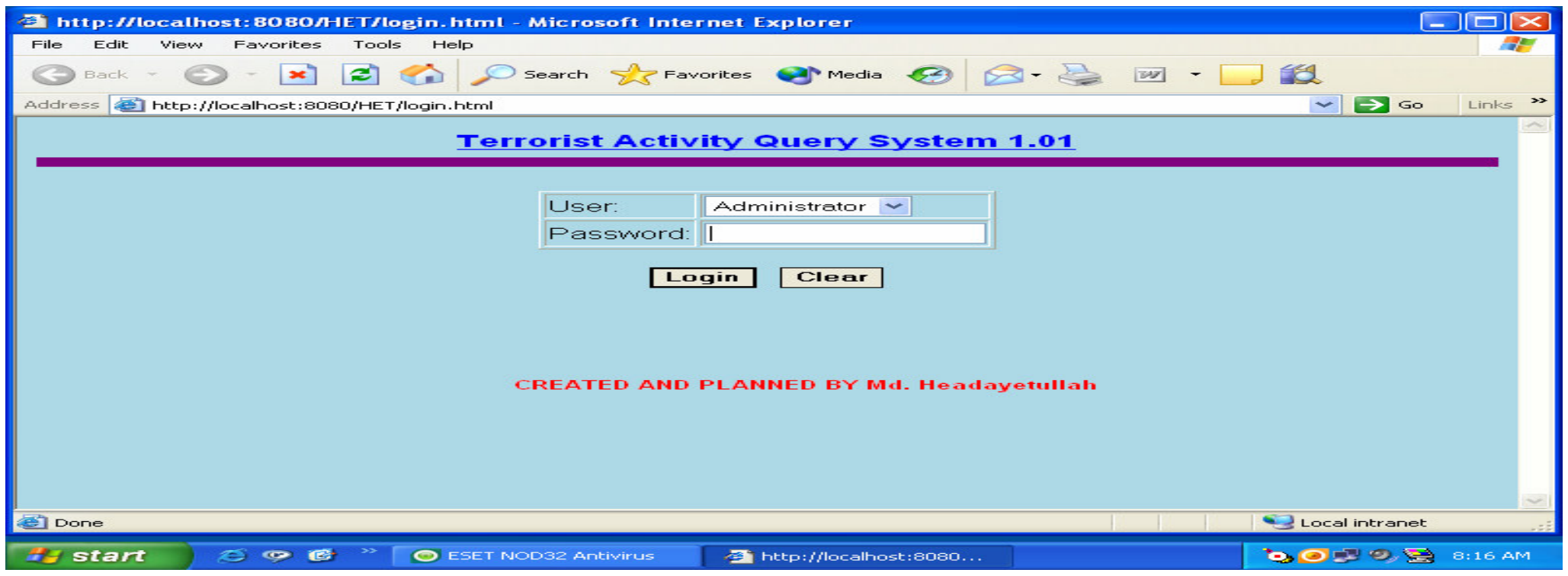

**Figure 3:** Snapshot of Administrator and Password verification

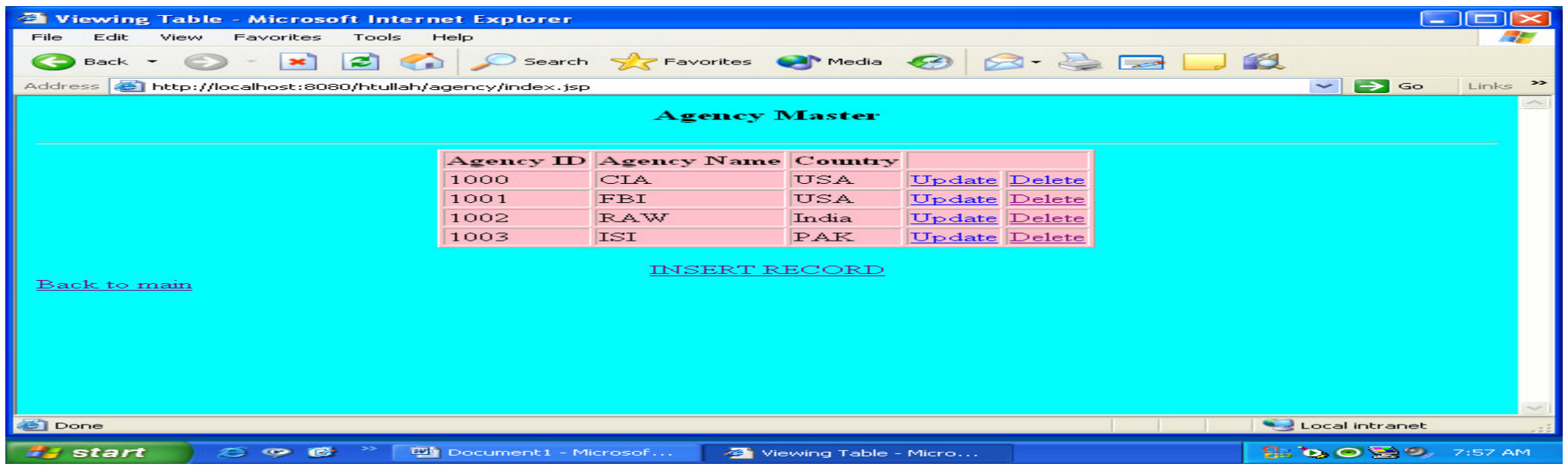

**Figure 4:** Snapshot of Agency Master

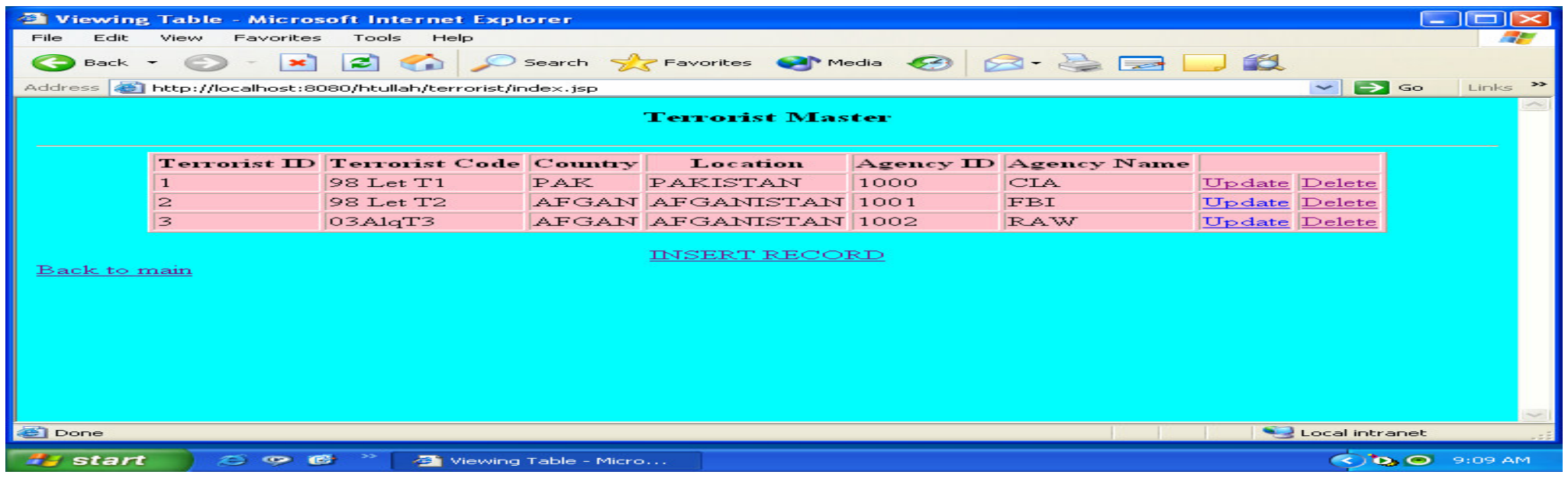

**Figure 5:** Snapshot of Terrorist Master

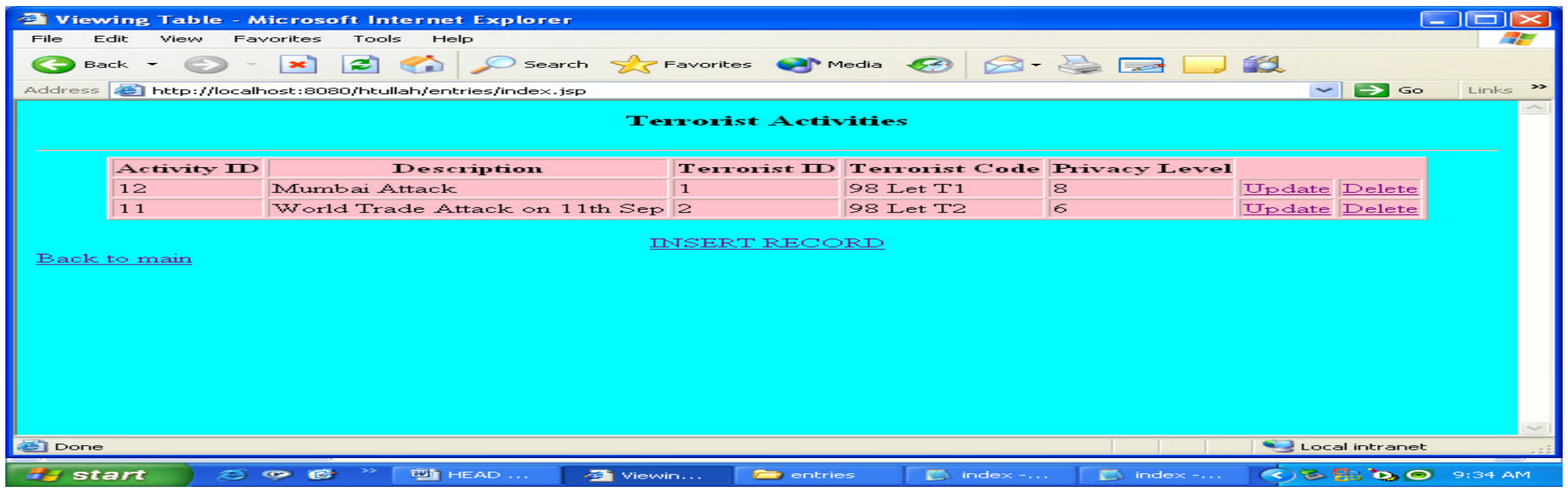

**Figure 6:** Snapshot of Terrorist Activities

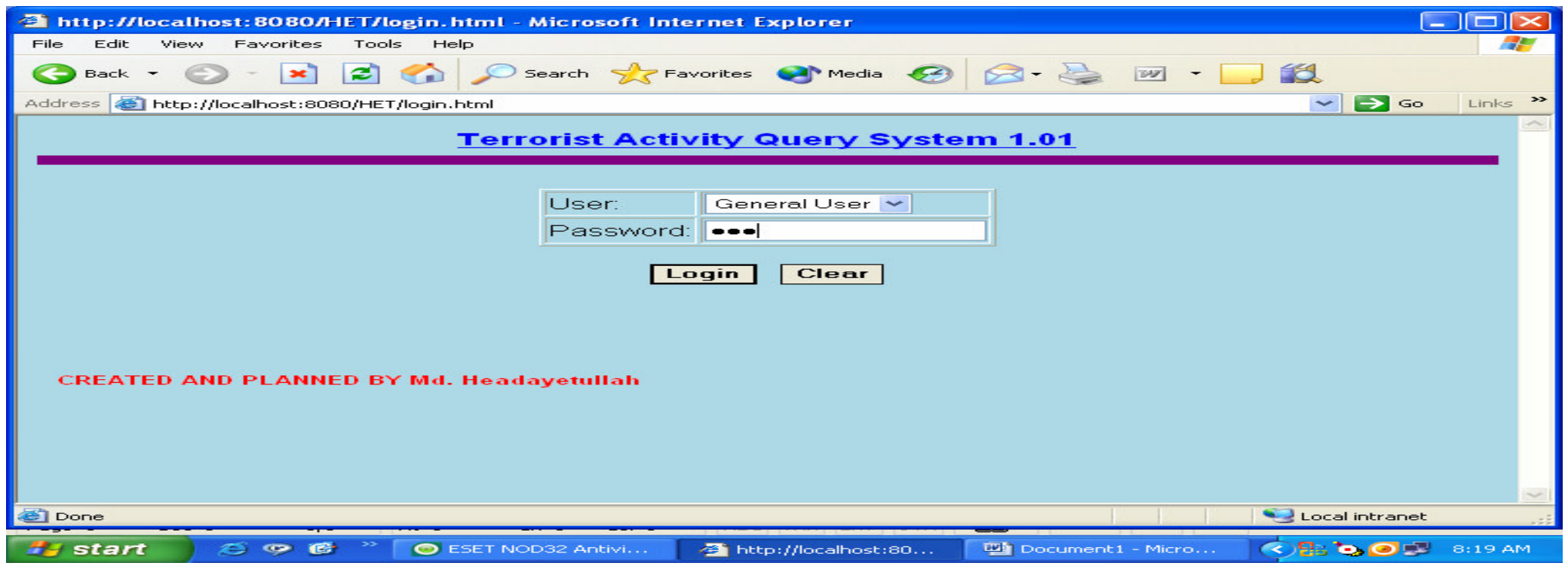

**Figure 7:** Snapshot of General User and Password

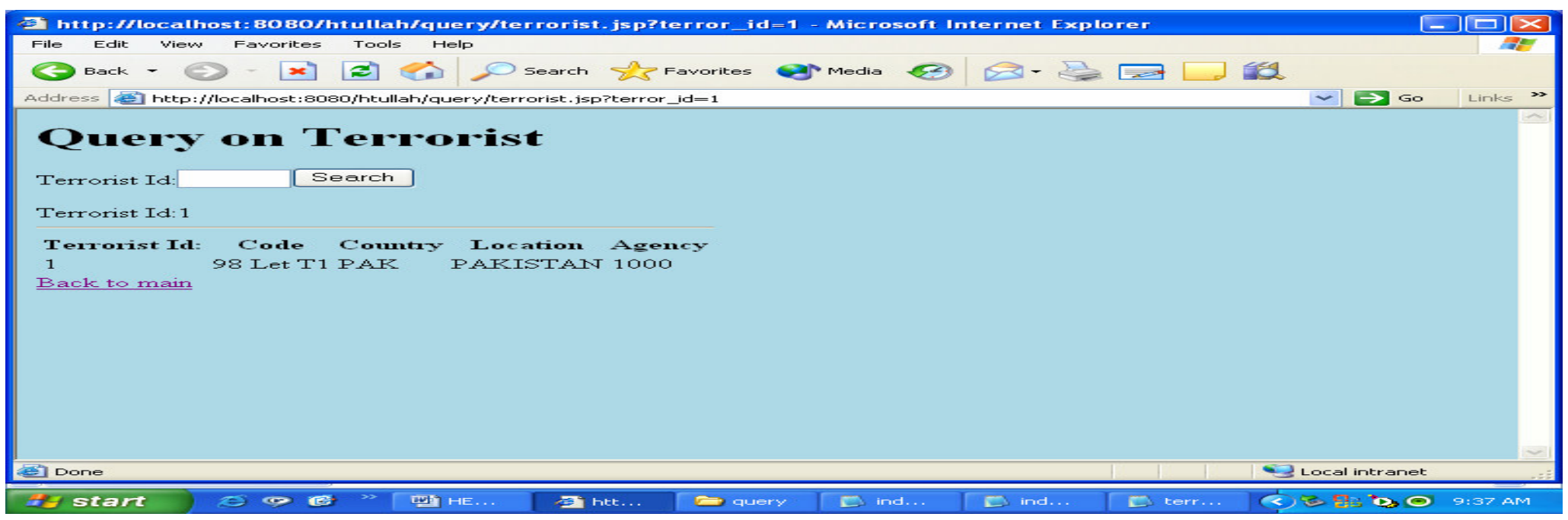

**Figure 8**: Snapshot of Query on Terrorist

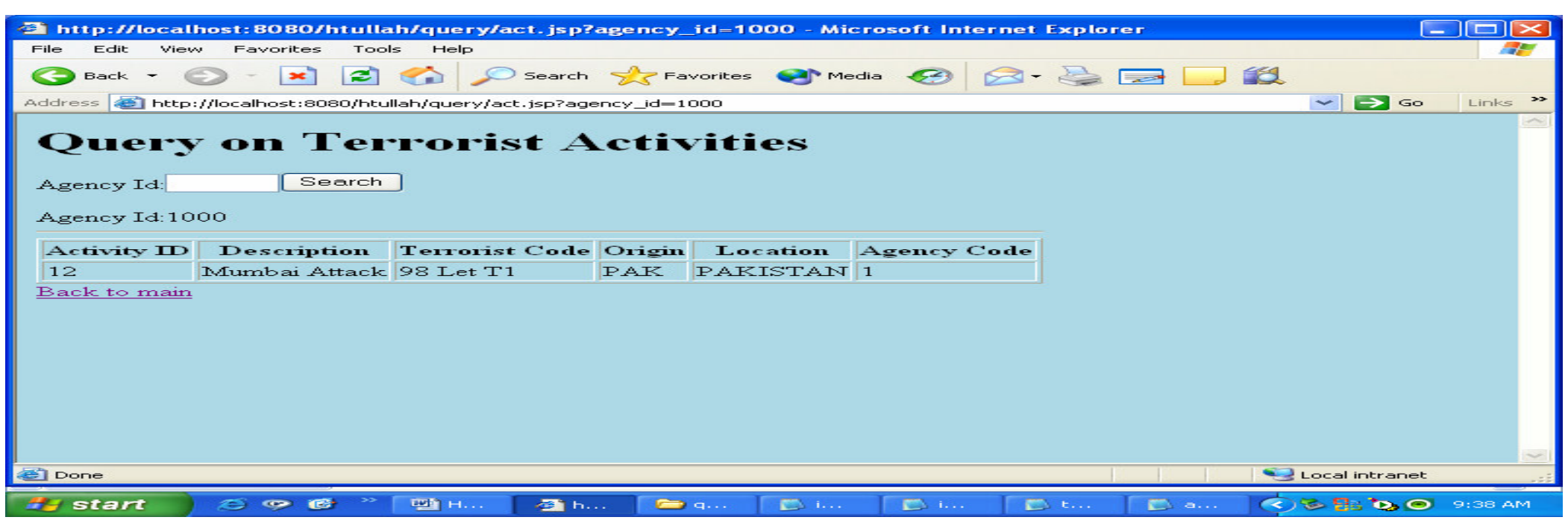

**Figure 9:** Snapshot of Query on terrorist activities

The experimental results of the obtainable trust-based information sharing protocol are presented in this section. The results acquired from experiments show that the offered protocol provides an effective and secure information sharing mechanism for communicating government intelligence agencies. The presented protocol is described as a three-way handshaking procedure to accomplish secure information sharing. The process started with a request for secret information about terrorists and their activities, by utilizing the techniques of hashing, a unique mapping function and public key cryptography. The target agency after a security verification replied with the suitable information on the basis of the trust level maintained with the source agency. The information shared will be a subset of the information available with the target agency based on the trust level. At the source agency, the legitimacy and confidentiality of the response is verified. Table 1 portrays the outcome obtained from the experimentation on the proposed trust-based security protocol using reproduction data.

**Table 1:** Results of Experimentation

| Source Agency | Target Agency | Terrorist Code | Information available with the Target agency | Trust-based Shared Information |
|---|---|---|---|---|
| CIA | FBI | 98LetT1 | {11,12,13,14,15,16,17,18,19,20} | {16,13,15,18,12,11,19,20,14} |
| ISI | CIA | 98LetT2 | {21,22,23,24,25,26,27,28,29,30} | {26,23,25,28,22,21,29,20,24} |
| RAW | CIA | 03AlqT3 | {31,32,33,34,35,36,37,38,39,40} | {37,34,38,35} |
| RAW | FBI | 06TalT4 | {41,42,43,44,45,46,47,48,49,50} | {49,42,46,44,48} |
| CIA | RAW | 98LetT5 | {51,52,53,54,55,56,57,58,59,60} | {56,53,55,58,52,51,59,60} |
| RAW | CIA | 06TalT6 | {61,62,63,64,65,66,67,68,69,70} | {69,62,66} |
| FBI | RAW | 98LetT7 | {71,72,73,74,75,76,77,78,79,80} | {76,73,75,78,72,71} |
| ISI | FBI | 03AlqT8 | {81,82,83,84,85,86,87,88,89,90} | {87,84,88,85,89} |
| CIA | FBI | 06TalT9 | {91,92,93,94,95,96,97,98,99,100} | {99,92,96,94} |
| ISI | FBI | 98LetT8 | {81,82,83,84,85,86,87,88,89,90} | {86,83,85,88,82,81,89,90} |

It is observable that the amount of information shared among communicating government intelligence agencies rely on the trust level maintained involving them. Within Table 1, the field “Information available in the target agency” gives inclusive safekeeping information available with the target intelligence agency concerning the terrorist and their distrustful actions, which has been composed over long periods of time and the field “Trust-based shared Information” consists of the information shared between the government intelligence agencies based on the trust level or privacy level, without harming their own national welfare. The experimental outcome reveal that the offered trust-based security protocol enables efficient and safe and sound information sharing on the basis of trust between worldwide government intelligence agencies without affect their own national interests. Therefore enhanced interoperability can be achieved between government intelligence agencies and other agencies based on trust protocol which will make more successful digital government.

## 5. CONCLUSION

Trust-based information exchange is an imperative trait of any digital government that needs to assure democratic values. The proposed trust- based protocol will aid to set up an enhance interoperability, concerning global government intelligence agency, local government intelligence agencies, public and private agencies by making use of top secret information sharing among them and which is the key significance issue to devise more successful digital government . Challenges in building a computational infrastructure for exchanging top secret information is difficult to solve and demand innovative motivation schemes. In this article, we have presented an original, expert and trust-based security protocol for confidential sharing of secret information amid government intelligence agencies. The designed trust-based security protocol has offered confidentiality, authentication, integrity, agency verification and a restricted privacy by utilizing public key infrastructure, MD5 Algorithm, a unique mapping function and predefined trust level respectively. The proposed protocol will enhance the interoperation level or trust level for interoperability and security in digital government by facilitating proficient and confidential sharing of top secret information. In short, the proposed work will improve the interoperability and security of digital government by making use of trust based protocol to share the top secret information among the government intelligence agencies to protect any threatening or unsocial activities. Thus, interoperability, trust based information sharing protocol and security are key issues of every modern Digital Government.

## Authors

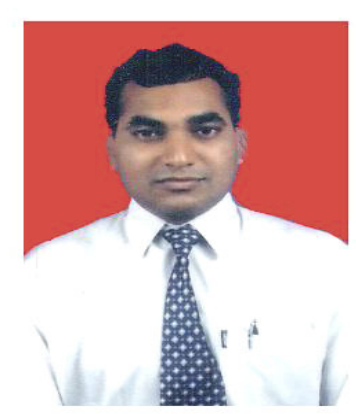

**Md.Headayetullah** received the Diploma in Computer Science & Engineering (DCSE) with 1st Class from Acharya Polytechnic, Bangalore, India and Bachelor of Engineering (B.E) degree with 1st Class from Yeshwantrao Chavan College of Engineering of Nagpur University, Nagpur, India in 2000 and 2003 respectively. He received second prize in state level for his best project in B.E degree. He received M.Tech degree with First Class with Honours from the Department of Computer Science & Engineering and Information Technology of Allahabad Agricultural Institute-Deemed University, Allahabd, India in 2005. He was the topper of the University in his M.Tech Degree. He is currently pursuing the PhD (Computer Sc. & Engineering) degree, working closely with Prof. (Dr.) G.K Pradhan and Prof. (Dr.) Sanjay Biswas in the Department of Computer Science and Engineering from Institute of Technical Education & Research (Faculty of Engineering) of Siksha O'Anusandhan University (SOAU), Bhubaneswar, India. He works in the field of E-Government, Digital Government, Networking, Internet Technology, Data Privacy, Cryptography, Information Security and Mobile Communication. He has authored four International Research Publication in Journal. He is currently working as an Assistant Professor in Computer Science & Engineering and Information Technology at Dr. B.C. Roy College of Engineering, Duragpur, West Bengal University of Technology, Kolkata, India.

**G.K. Pradhan** received the PhD degree from Indian Institute of Technology (IIT) Kanpur, India. He served as a lecturer, Assistant Professor and Associate Professor in several Institutes in India. Dr. Pradhan is currently working as a Professor in Computer Science & Engineering and Information Technology at Institute of Technical Education & Research (Faculty of Engineering) of Siksha O Anusandhan University (SOAU), Bhubaneswar, India. He is working as a Chair person of Doctoral Scrutiny Committee (DSC) of Institute of Technical Education & Research (Faculty of Engineering) of Siksha O Anusandhan University, Bhubaneswar, India. Dr. Pradhan serves as a research supervisor for PhD degree in the field of Computer Science and Engineering and mathematics. He has authored several books and had more than 25 research publication in Journal. His filed of interests are Software Engineering, E-Commerce, E-Business Application, Digital Government, Internet Technology and Mobile Communication.